\documentclass[10pt, technote]{IEEEtran}

\usepackage{capjelcommands}

\newcommand{\OT}{\left(\boldsymbol{r}, \omega\right)} 
\newcommand{\RT}{\left(\boldsymbol{r}, t\right)} 

\begin{document}
\title{Various Interpretations of the Stored and the Radiated Energy Density}
\author{Miloslav~Capek,~\IEEEmembership{Member,~IEEE,}~and~Lukas~Jelinek
\thanks{Manuscript received XXX, 2015; revised XXX, 2015.
This work was supported by the Czech Science Foundation under project No.~15-10280Y}
\thanks{M.~Capek and L.~Jelinek are with the Department of Electromagnetic Field, Faculty of Electrical Engineering, Czech Technical University in Prague, Technicka 2, 16627, Prague, Czech Republic
(e-mail: \mbox{miloslav.capek@fel.cvut.cz}, \mbox{lukas.jelinek@fel.cvut.cz}).}}

\markboth{Journal of \LaTeX\ Class Files,~Vol.~XX, No.~XX, xxx~20XX}
{Capek, Jelinek: On the different looks on the stored and radiated energy of a radiating system} 
\maketitle

\begin{abstract}
Three contradictory but state-of-the-art concepts for defining and evaluating stored electromagnetic energy are treated in this communication, and are collated with the widely accepted definition of stored energy, which is the total energy minus the radiated energy. All three concepts are compared, and the results are discussed on an example of a dominant spherical mode, which is known to yield dissimilar results for the concepts dealt with here. It is shown that various definitions of stored energy density immanently imply diverse meanings of the term ``radiation''.
\end{abstract}


\section{Introduction}
\label{Intro}
\IEEEPARstart{T}{he} evaluation of stored electromagnetic energy and its density is one of the old but as yet unsolved problems of classical electromagnetism. This is true despite its straightforward and generally accepted definition: stored electromagnetic energy is that part of the total electromagnetic energy that is, in comparison with the radiated energy, bound to the sources of the field, being unable to escape towards infinity. In other words, the stored electromagnetic energy is the difference between the total and the radiated electromagnetic energies, representing a ``rest mass'' seen by the force exerted by the source. 

In the case of a static field and a quasi-static field, the evaluation of the stored energy is immediate, as it is just equal to the total energy \cite{Jackson_ClassicalElectrodynamics}, the physical interpretation of which is directly inferred from Poynting's theorem \cite{Jackson_ClassicalElectrodynamics}. The problem arises for fields generated by general radiators. One of the core problems is that within the time harmonic steady state, the total energy is infinite \cite{Jackson_ClassicalElectrodynamics}. This infinite energy is contained in the radiation field, and the goal of the evaluation of the stored energy is to subtract this infinite radiation energy. The second problem is that Poynting's theorem gives us no clue of what radiation energy really is. One can only rely on some general properties like the positive semi-definiteness of the energy, and the fact that the far field of the radiator placed in the lossless media carries solely radiation energy \cite{Schwinger_ClassicalElectrodynamics}. 
 
Probably the first treatment of stored electromagnetic energy dates back to the work of Bateman \cite{Bateman_TheMathematicalAnalysisOfElectricalAndOpticalWaveMotion}, who pointed out that the electromagnetic energy in a vacuum does not in general move with the speed of light in a vacuum, being slowed down by a kind of ``rest mass". This ``rest mass" vanishes only in the case of pure radiation fields, i.e. only at an infinite distance from a finite source. This work has however been forgotten and it was not until recent years that it was recovered and generalized by Kaiser \cite{Kaiser_ElectromagneticInertiaReactiveEnergyAndEnergy}, into the form of stored electromagnetic energy density.

In parallel, the problem of stored energy has also been extensively studied in the community of electrical engineering, mostly in connection with antennas. In particular, antenna designers commonly aim at the lowest energy storage in order to maximize the radiation efficiency of an antenna, and for this purpose one actually encounters the problem of stored energy evaluation \cite{VolakisChenFujimoto_SmallAntennas}. To that point, Chu \cite{Chu_PhysicalLimitationsOfOmniDirectAntennas} proposed a circuit equivalent of the spherical modes, and with its help subtracted the radiation energy and established fundamental lower bounds of the radiation quality factor. His method has been generalized by several works of Thal \cite{Thal_RadiationQandGainOfTMandTEsourcesInPhaseDelayedRotatedConfigurations, Thal_PolarizationGainAndQforSmallAntennas, Thal_QboundsForArbitrarySmallAntennas}. Radiation energy subtraction has also been attempted directly on the field level. Among the most prominent works we mention \cite{CollinRotchild_EvaluationOfAntennaQ, Collin_MinimumQofSmallAntennas, Rhodes_OnTheStoredEnergyOfPlanarApertures, Rhodes_AReactanceTheorem, Vandenbosch_ReactiveEnergiesImpedanceAndQFactorOfRadiatingStructures}, and also \cite{MikkiAntar_ATheoryOfAntennaElectromagneticearField1}, in which subtraction is taken from another point of view, claiming that separation of total energy into stored energy and radiated energy is unphysical, as it cannot be derived from Maxwell's equations.

The works of Bateman \cite{Bateman_TheMathematicalAnalysisOfElectricalAndOpticalWaveMotion}, Kaiser \cite{Kaiser_ElectromagneticInertiaReactiveEnergyAndEnergy}, and Collin \cite{Collin_MinimumQofSmallAntennas} attempted to define stored energy locally via its density, while the works of Rhodes \cite{Rhodes_ObservableStoredEnergiesOfElectromagneticSystems}, Yaghjian and Best \cite{YaghjianBest_ImpedanceBandwidthAndQOfAntennas}, Vandenbosch \cite{Vandenbosch_ReactiveEnergiesImpedanceAndQFactorOfRadiatingStructures}, and Gustafsson and Jonsson \cite{Gustaffson_StoredElectromagneticEnergy_PIER} operated solely with stored energy as a whole. The local approach however offers certain benefits: first, radiation energy extraction is carried out at every space-time point, avoiding cumbersome operations with ill-defined infinite space integrations \cite{Vandenbosch_ReactiveEnergiesImpedanceAndQFactorOfRadiatingStructures}, and second, the definition via the density is also more physical, as within classical relativistic theory all laws should be formulated strictly locally.

The concepts mentioned above yield similar values of stored energy for common radiators. However, there exist specific cases for which the methods are profoundly different. One such case is the dominant TE spherical mode, which has been shown to lead to apparently incorrect negative stored energies within some evaluation schemes \cite{GustafssonCismasuJonsson_PhysicalBoundsAndOptimalCurrentsOnAntennas_TAP}.

This paper has two main purposes: to compare different interpretations of radiated energy, and to recall the possibility of a local definition of stored energy, which, despite its appealing properties, has not found its place within the antenna community.

The paper is organized as follows. The necessary definitions and nomenclature are introduced in Sec.~\ref{Sec_Defs}. Sec.~\ref{Sec_Meth} recalls several radiation energy extraction techniques. The techniques are compared on an example of the dominant spherical mode in Sec.~\ref{Sec_Example}. The results are discussed in Sec.~\ref{Sec_Disc}. Conclusions are drawn in Sec.~\ref{Sec_Concl}.

\section{Definitions}
\label{Sec_Defs}

In this paper, we will strictly omit dispersive media, in which the concept of stored energy is problematic, even without radiation \cite{LandauLifshitzPitaevskii_ElectrodynamicsOfContinuousMedia}. Furthermore, as is common in the theory of electromagnetic radiators, this paper will deal with time harmonic fields represented by field phasors \mbox{$\V{F} \left(\V{r},\omega\right)$} at angular frequency $\omega$, which relates to the time domain quantities as \mbox{$\boldsymbol{\mathcal{F}} \left(\boldsymbol{r},t\right) = \RE\left\{\V{F}\left(\V{r},\omega\right)\mathrm{exp}(\J\omega t)\right\}$}. We will also use cycle mean averages, which are defined as \mbox{$\langle f \left( t \right) \rangle = \left(1 / T\right) \int_{t_0}^{t_0+T} f \left(t\right)\D{t}$}, with \mbox{$T = 2 \pi / \omega$}, and are widely employed in the case of power quantities, where \mbox{$\langle \boldsymbol{\mathcal{F}} \left(\V{r},t\right) \cdot \boldsymbol{\mathcal{G}} \left(\V{r},t\right) \rangle = \left(1 / 2\right) \RE\left\{ \V{F} \left(\V{r},\omega\right) \cdot \V{G}^\ast \left(\V{r},\omega\right) \right\}$} within the time harmonic domain \cite{Harrington_TimeHarmonicElmagField}, and in which symbol $\ast$ denotes complex conjugate. Note, however, that the results detailed below and summarized in Table~\ref{ref_Table1} hold for a general time domain field.

For the purpose of comparing various concepts of stored energy, we will advantageously use a dimensionless quality factor, defined as
\BE
\label{Eq1}
Q = \frac{\omega W_\mathrm{sto}}{P_\mathrm{r}},
\EE
where 
\BE
\label{Eq2}
W_\mathrm{sto} = \int\limits_{V} \langle w_\mathrm{sto} \left(\V{r},t\right) \rangle \D{V}
\EE
is the total cycle mean stored energy, defined via its density \mbox{$w_\mathrm{sto} \left(\V{r},t\right)$}, and where 
\BE
\label{Eq3}
\begin{split}
P_\mathrm{r} &= \oiint\limits_S \langle \boldsymbol{\mathcal{E}} \left(\V{r},t\right) \times \boldsymbol{\mathcal{H}} \left(\V{r},t\right) \rangle \cdot \D{\V{S}} \\
&= \frac{1}{2} \oiint\limits_S \RE\left\{ \V{E} \left(\V{r},\omega\right) \times \V{H}^\ast \left(\V{r},\omega\right) \right\}\cdot\D{\V{S}}
\end{split}
\EE
is the cycle mean radiated power \cite{Jackson_ClassicalElectrodynamics}.

\section{Various Definitions of Stored Energy Density and Radiated Energy Density}
\label{Sec_Meth}

This section briefly reviews three major definitions of stored electromagnetic energy density used in the literature. The three concepts are summarized in Table~\ref{ref_Table1}, and their general properties are discussed in the following section.
\begin{table*}[t!]
\begin{center}
\caption{Various concepts of extraction and subtraction of radiated energy.}
{\tabulinesep=1.2mm
\label{ref_Table1}
\begin{tabu}{|c||c|c|}
\hline Concept & $\mathcal{W}_\mathrm{sto}\left(\V{r}, t\right)$ & $\mathcal{W}_\mathrm{rad}\left(\V{r}, t\right)$ \\ 
\hline
\hline Collin and Rothschild \cite{CollinRotchild_EvaluationOfAntennaQ} & $\mathcal{U}\left(\V{r}, t\right) - \mathcal{W}_\mathrm{rad}\left(\V{r}, t\right)$ & $\displaystyle\frac{\boldsymbol{\mathcal{S}}\left(\V{r}, t\right)}{c_0} \cdot \UV{n}$ \\ 
\hline Kaiser \cite{Kaiser_ElectromagneticInertiaReactiveEnergyAndEnergy}, Bateman \cite{Bateman_TheMathematicalAnalysisOfElectricalAndOpticalWaveMotion} & $\sqrt{\mathcal{U}^2\left(\V{r}, t\right) - \mathcal{W}_\mathrm{rad}^2\left(\V{r}, t\right)}$ & $\displaystyle\left\|\frac{\boldsymbol{\mathcal{S}}\left(\V{r}, t\right)}{c_0}\right\|$ \\ 
\hline Rhodes \cite{Rhodes_ObservableStoredEnergiesOfElectromagneticSystems}, Yaghjian and Best \cite{YaghjianBest_ImpedanceBandwidthAndQOfAntennas}, Vandenbosch \cite{Vandenbosch_ReactiveEnergiesImpedanceAndQFactorOfRadiatingStructures}, Gustafsson and Jonsson \cite{Gustaffson_StoredElectromagneticEnergy_PIER} & $\mathcal{U}\left(\V{r}, t\right) - \mathcal{W}_\mathrm{rad}\left(\V{r}, t\right)$ &
$\displaystyle\epsilon_0\frac{\left\| \boldsymbol{\mathcal{F}}\left(\V{r}, t\right) \right\|^2}{r^2}$ \\ 
\hline 
\hline 
\multicolumn{3}{|c|}{$\mathcal{U}\left(\V{r}, t\right) = \frac{1}{2} \epsilon_0 \|\boldsymbol{\mathcal{E}}\left(\V{r}, t\right)\|^2 + \frac{1}{2} \mu_0 \|\boldsymbol{\mathcal{H}}\left(\V{r}, t\right)\|^2$, $\quad\boldsymbol{\mathcal{S}}\left(\V{r}, t\right) = \boldsymbol{\mathcal{E}}\left(\V{r}, t\right)\times\boldsymbol{\mathcal{H}}\left(\V{r}, t\right)$, $\quad\boldsymbol{\mathcal{F}} \left(\V{r}, t\right) = \lim\limits_{r\rightarrow\infty} \left(r \boldsymbol{\mathcal{E}} \left(\boldsymbol{r},t + \frac{r}{c_0}\right)\right)$}  \\ 
\hline 
\end{tabu}}
\end{center}
\end{table*}

\subsection{The Concept of Collin-Rothschild}
\label{Sec_DefsA}

The classical scheme for radiation energy extraction was defined by Collin and Rothschild in \cite{CollinRotchild_EvaluationOfAntennaQ}, and was later refined in \cite{Collin_MinimumQofSmallAntennas} into the form of energy density exposed in the first column of Table~\ref{ref_Table1}. The corresponding cycle mean energy density for a time harmonic field can also easily be written  as
\BE
\label{Eq4}
\begin{split}
\langle w_\mathrm{sto}^\mathrm{CR} \OT \rangle = &\frac{1}{4} \left( \epsilon \|\V{E} \OT\|^2 + \mu \|\V{H}\OT\|^2\right) \\
&- \frac{1}{2 c_0} \RE\left\{ \V{E} \OT \times \V{H}^\ast \OT \right\} \cdot \UV{n},
\end{split}
\EE
where $\V{n}_0$ is the normal to the far field wave-front. 

This radiation energy extraction is the most common method used in the literature \cite{VolakisChenFujimoto_SmallAntennas}, despite its immediate deficiency of using $\V{n}_0$ as the direction of the power flow. This poses no problem for specific geometries (pure modes in separable coordinate systems), but cannot suffice in general. Any general radiator will clearly in its near field emit radiation in directions different from the far field.

For the sake of comparison, the quantity
\BE
\label{Eq5}
Q^\mathrm{RC} = \frac{\omega W_\mathrm{sto}^\mathrm{RC}}{P_\mathrm{r}}
\EE
is defined as advantageously normalized total stored energy within this concept.

\subsection{The Concept of Kaiser-Bateman}
\label{Sec_DefsB}

A very interesting way of evaluating stored energy was proposed by Kaiser \cite{Kaiser_ElectromagneticInertiaReactiveEnergyAndEnergy}, generalizing the previous work of Bateman \cite{Bateman_TheMathematicalAnalysisOfElectricalAndOpticalWaveMotion}. Within this concept, the relativistic energy-momentum relation \cite{Jackson_ClassicalElectrodynamics} is used to define the stored energy density exposed in the second row of Table~\ref{ref_Table1}.

To the best of the authors' knowledge, this is the first time that stored energy density has been defined strictly locally with no reference to the position of the sources, i.e. radiation energy extraction is carried out locally at every point of the space-time. 

Using time harmonic fields, the cycle mean stored energy within this concept can be written explicitly as
\BE
\label{Eq6}
\langle w_\mathrm{sto}^\mathrm{KB} \RT \rangle = \bigg\langle\sqrt{U^2 \RT - W_\mathrm{rad}^2 \RT}\bigg\rangle,
\EE
in which
\BE
\label{Eq7}
\begin{split}
U^2 \RT -& W_\mathrm{rad}^2 \RT = \frac{1}{16} \Big( \epsilon \left\|\V{E}\OT\right\|^2 + \mu \left\|\V{H}\OT \right\|^2 \\
& + \frac{1}{2} \RE\Big\{ \big(\epsilon \V{E} \OT \cdot \V{E}\OT \\
& + \mu \V{H}\OT \cdot \V{H}\OT \big) \mathrm{e}^{2 \J \omega t} \Big\} \Big)^2 \\
& - \frac{1}{4} \bigg\| \frac{1}{c_0} \big( \RE\left\{ \V{E}\OT \times \V{H}^\ast\OT \right\} \\
& + \RE\left\{ \V{E}\OT \times \V{H}\OT \mathrm{e}^{2 \J \omega t} \right\} \big) \bigg\|^2
\end{split}
\EE
with the corresponding quality factor 
\BE
\label{Eq8}
Q^\mathrm{KB} = \frac{\omega W_\mathrm{sto}^\mathrm{KB}}{P_\mathrm{r}}.
\EE

Two crucial differences immediately appear when comparing (\ref{Eq4}) and (\ref{Eq6}). First, in (\ref{Eq6}) the entire power flow is subtracted from the field energy, while in (\ref{Eq4}) only the power flow along the direction of wave-front at infinity is subtracted. Second, in (\ref{Eq6}) the subtraction in done is squares, while in (\ref{Eq4}) the subtraction is direct. As a result of the squared subtraction in (\ref{Eq7}), the cycle mean cannot be simply performed a priori.

With respect to this concept of Kaiser-Bateman, it is worth mentioning a situation in which the general time domain definition, see Table~\ref{ref_Table1}, can be greatly simplified \cite{Kaiser_ElectromagneticInertiaReactiveEnergyAndEnergy}. This happens in the case when \mbox{$\boldsymbol{\mathcal{E}} \RT \cdot \boldsymbol{\mathcal{H}} \RT = 0$}, which is the case for the example in this paper, and it is not rare even in realistic situations (at least in an approximate sense). Under this specific condition, the definition from the second row of Table~\ref{ref_Table1} can be rewritten as
\BE
\label{Eq9}
w_\mathrm{sto}^\mathrm{KB} \RT = \frac{1}{2} \left|\epsilon \left\| \boldsymbol{\mathcal{E}}\RT \right\|^2 - \mu \left\| \boldsymbol{\mathcal{H}}\RT \right\|^2\right|
\EE
which is a rather curious form. Not being the absolute value, (\ref{Eq9}) would correspond to the energy excess appearing in the complex Poynting's theorem \cite{Jackson_ClassicalElectrodynamics}. The absolute value, however, makes it (according to Kaiser and Bateman) the stored energy density.

\subsection{The Concept of Rhodes}
\label{Sec_DefsC}
Another well-established scheme for stored energy evaluation was used by Rhodes \cite{Rhodes_ObservableStoredEnergiesOfElectromagneticSystems}, was generalized by Yaghjian \cite{YaghjianBest_ImpedanceBandwidthAndQOfAntennas} and was later reworked into the source concept by Vandenbosh \cite{Vandenbosch_ReactiveEnergiesImpedanceAndQFactorOfRadiatingStructures} for radiators of arbitrary shape. The definition of stored energy density within this concept is exposed in the third row of Table~\ref{ref_Table1}. The formula can also easily be rewritten for time harmonic fields as
\BE
\label{Eq10}
\begin{split}
\langle w_\mathrm{sto}^\mathrm{Rh} &\RT \rangle \\
&= \frac{1}{4} \Bigg\langle \epsilon \left\| \V{E} \OT \right\|^2 + \mu \left\| \V{H} \OT \right\|^2 - 2 \epsilon \frac{\left\|\V{F} \right\|^2}{r^2} \Bigg\rangle,
\end{split}
\EE
in which \mbox{$\V{F} = \lim\limits_{r\rightarrow\infty} \left( r \V{E} \, \exp\left(\J k r\right) \right)$}.
The corresponding normalization is defined as
\BE
\label{Eq11}
Q^\mathrm{Rh} = \frac{\omega W_\mathrm{sto}^\mathrm{Rh}}{P_\mathrm{r}}.
\EE

\section{Stored Energy and Its Density For the Dominant Spherical TE Mode}
\label{Sec_Example}

\begin{figure}
\centering
\includegraphics[width=8.9cm]{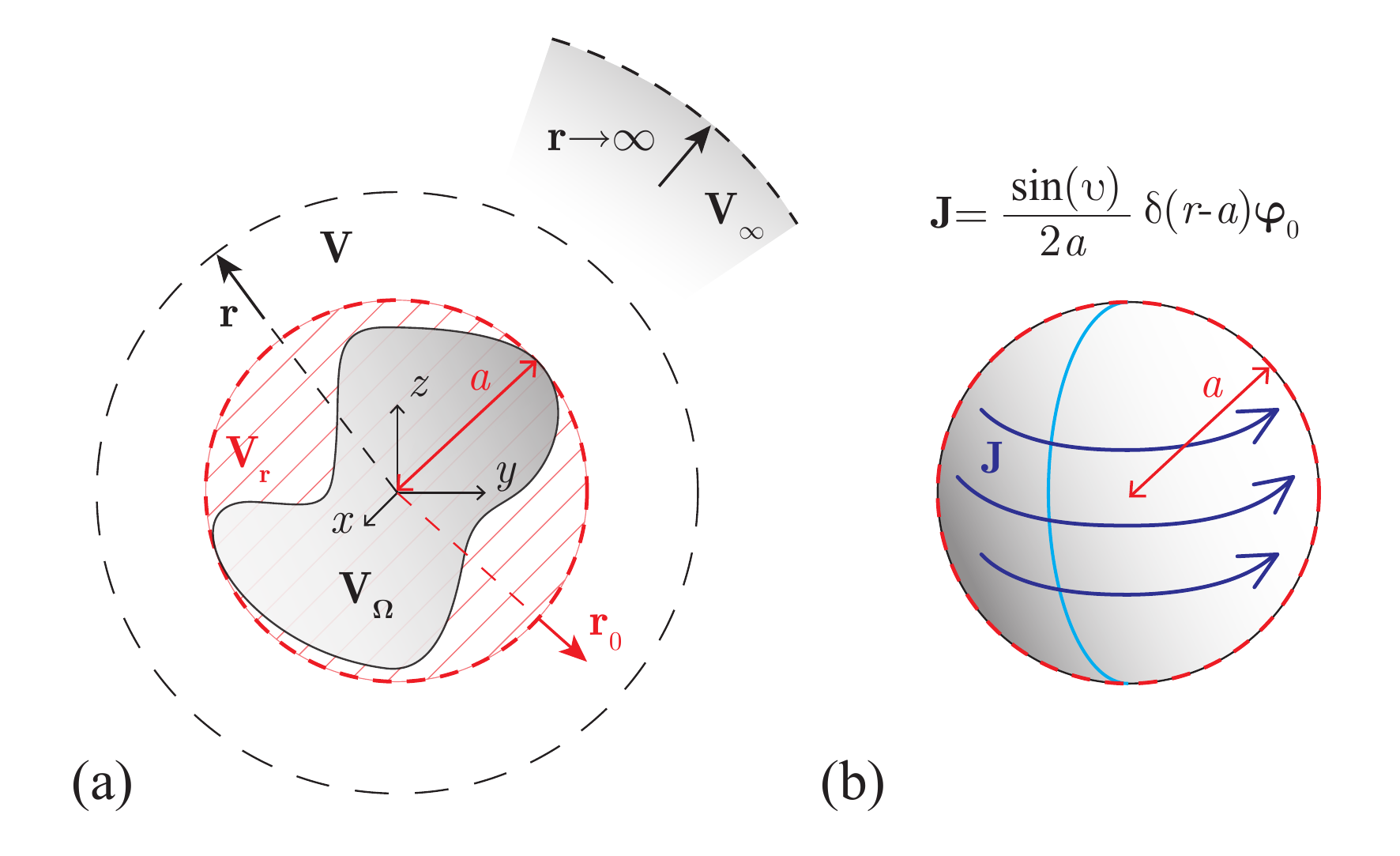}
\caption{Sketch of the coordinate system that is used throughout the paper (a) and a sketch of the $\mathrm{TE}_{10}$ current on a spherical shell of radius $a$. (b) The input current is normalized to $I_0 = 1$\,A with respect to the light blue contour.}
\label{fig_figure1}
\end{figure}

The dominant spherical TE mode is defined as the field generated by the current density \cite{Jackson_ClassicalElectrodynamics}
\BE
\label{Eq12}
\V{J} \left(\vartheta, a\right) = \frac{\sin \left(\vartheta\right)}{2 a} \delta\left(r - a\right) \boldsymbol{\varphi}_0,
\EE
flowing on a spherical shell of radius $a$, see Fig.~\ref{fig_figure1}, and it provides interesting testing grounds for stored energy evaluation \cite{Gustaffson_StoredElectromagneticEnergy_PIER}. 
The fields generated by this source read \cite{Stratton_ElectromagneticTheory} for $r < a$
\begin{subequations}
\begin{align}
\label{Eq13A}
\V{E} =& - \frac{\mu\omega}{2} ka\, \mathrm{h}_1^{(2)} \left(ka \right) \mathrm{j}_1 \left(kr \right) \sin\left(\vartheta\right) \boldsymbol{\varphi}_0, \\
\label{Eq13B}
\V{H} =& \J k \frac{ka\, \mathrm{h}_1^{(2)} \left(ka\right)}{kr} \Big( -\mathrm{j}_1 \left(ka\right) \cos\left(\vartheta\right)\UV{r} \nonumber\\
&+ \frac{kr \, \mathrm{j}_0 \left(kr\right) - \mathrm{j}_1 \left(kr\right)}{2} \sin\left(\vartheta\right) \boldsymbol{\vartheta}_0 \Big),
\end{align}
\end{subequations}
and for $r \ge a$
\begin{subequations}
\begin{align}
\label{Eq14A}
\V{E} = &- \frac{\mu\omega}{2} ka\, \mathrm{j}_1 \left(ka \right) \mathrm{h}_1^{(2)} \left(kr \right) \sin\left(\vartheta\right) \boldsymbol{\varphi}_0, \\
\label{Eq14B}
\V{H} = & \J k \frac{ka\, \mathrm{j}_1 \left(ka\right)}{kr} \Big( -\mathrm{h}_1^{(2)} \left(ka\right) \cos\left(\vartheta\right)\UV{r} \nonumber\\
&+ \frac{kr \, \mathrm{h}_0^{(2)} \left(kr\right) - \mathrm{h}_1^{(2)} \left(kr\right)}{2} \sin\left(\vartheta\right) \boldsymbol{\vartheta}_0 \Big),
\end{align}
\end{subequations}
where $k = \omega / c_0$ is the free space wave-number and $c_0$ is the speed of light.

The total energy of the fields $W^\mathrm{tot}$ (in the form of the corresponding quality factor) can be evaluated in a straightforward manner as
\BE
\label{Eq15}
Q^\mathrm{tot} = \frac{\omega W^\mathrm{tot}}{P_\mathrm{r}} = Q_\mathrm{int}^\mathrm{tot} + Q_\mathrm{ext}^\mathrm{tot}
\EE
with
\begin{subequations}
\begin{align}
\label{Eq16A}
Q_\mathrm{int}^\mathrm{tot} &= \frac{\omega W_\mathrm{int}^\mathrm{tot}}{P_\mathrm{r}} = \int\limits_0^{ka} Q_\mathrm{int}^\mathrm{tot} \left(kr\right)\D{kr} \nonumber\\
&= \frac{1}{2} \frac{\left| h_1^{(2)} \left(ka\right) \right|^2}{j_1^2 \left(ka\right)} \int\limits_0^{ka} \Big( (kr)^2 \left|j_1\left(kr\right)\right|^2 + 2 \left|j_1 \left(kr\right)\right|^2 \nonumber\\ 
&\quad+ \left| kr \, j_0 \left( kr \right) - j_1 \left(kr\right) \right|^2 \Big)\D{kr}, \\
\label{Eq16B}
Q_\mathrm{ext}^\mathrm{tot} &= \frac{\omega W_\mathrm{ext}^\mathrm{tot}}{P_\mathrm{r}} = \int\limits_{ka}^{\infty} Q_\mathrm{ext}^\mathrm{tot} \left(kr\right)\D{kr} \nonumber\\
&= \frac{1}{2} \int\limits_{ka}^{\infty} \Big( (kr)^2 \left|h_1^{(2)}\left(kr\right)\right|^2 + 2 \left|h_1^{(2)} \left(kr\right)\right|^2 \nonumber\\ 
&\quad+ \left| kr \, h_0^{(2)} \left( kr \right) - h_1^{(2)} \left(kr\right) \right|^2 \Big)\D{kr},
\end{align}
\end{subequations}
where subscript ``int'' denotes energies for $r < a$ and subscript ``ext'' denotes the energies for $r \geq a$. The integration in (\ref{Eq16A})--(\ref{Eq16B}) can be carried out analytically (though it will lead to \mbox{$Q_\mathrm{ext}^\mathrm{tot} \rightarrow \infty$}), but we rather leave out the possibility to study the radial energy density represented by $Q \left(kr\right)$.

\begin{figure}
\centering
\includegraphics[width=8.9cm]{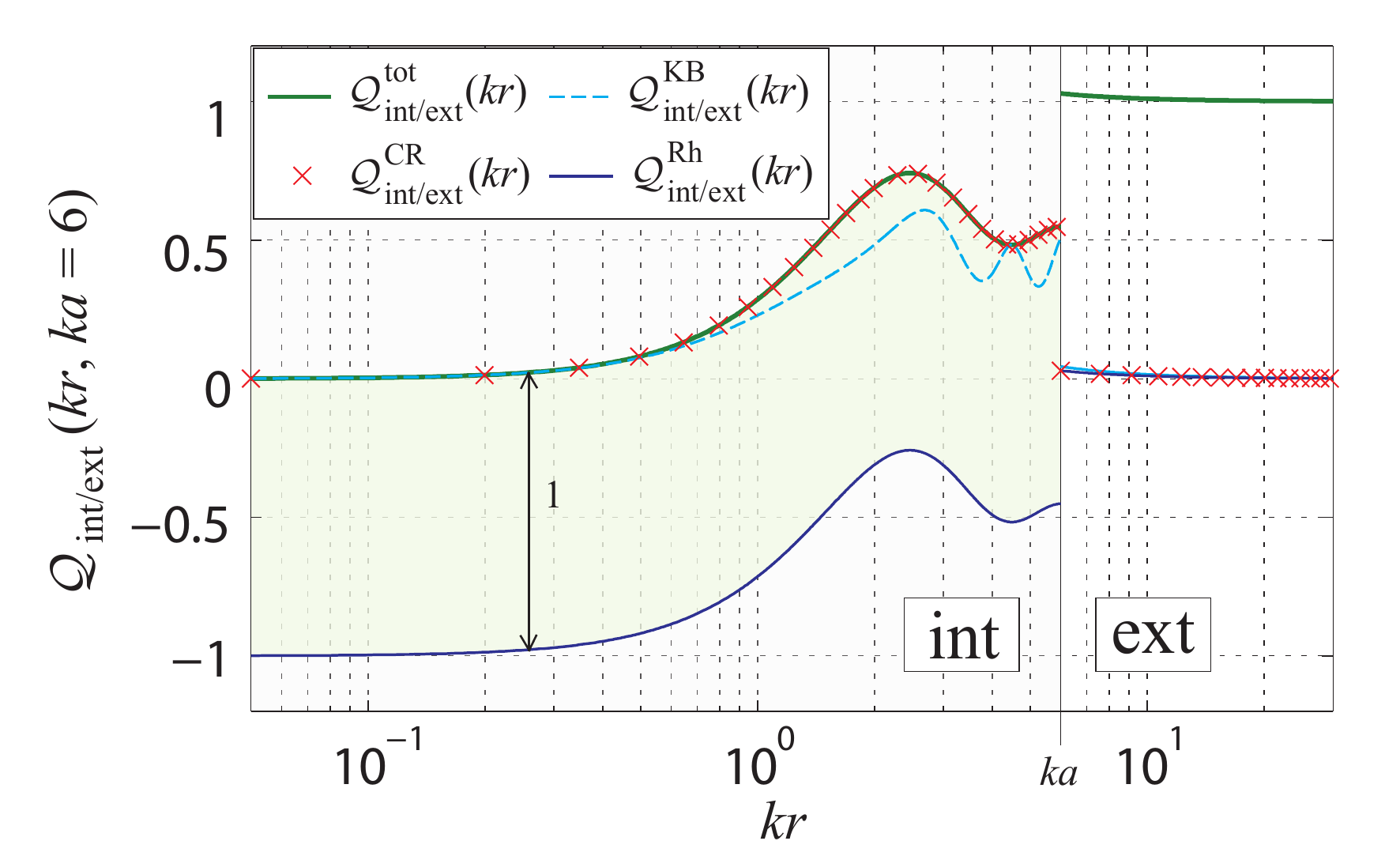}
\caption{The normalized total energy density and the normalized stored energy density of the spherical TE$_{10}$ mode for $ka = 6$. Several approaches to obtain the stored energy density are depicted. The energy jump is given by the presence of the current shell at $r=a$. The surface of the shaded area is equal to $ka$, which is in exact correspondence with \cite{Gustaffson_StoredElectromagneticEnergy_PIER}.}
\label{fig_figure2}
\end{figure}

\begin{figure}
\centering
\includegraphics[width=8.9cm]{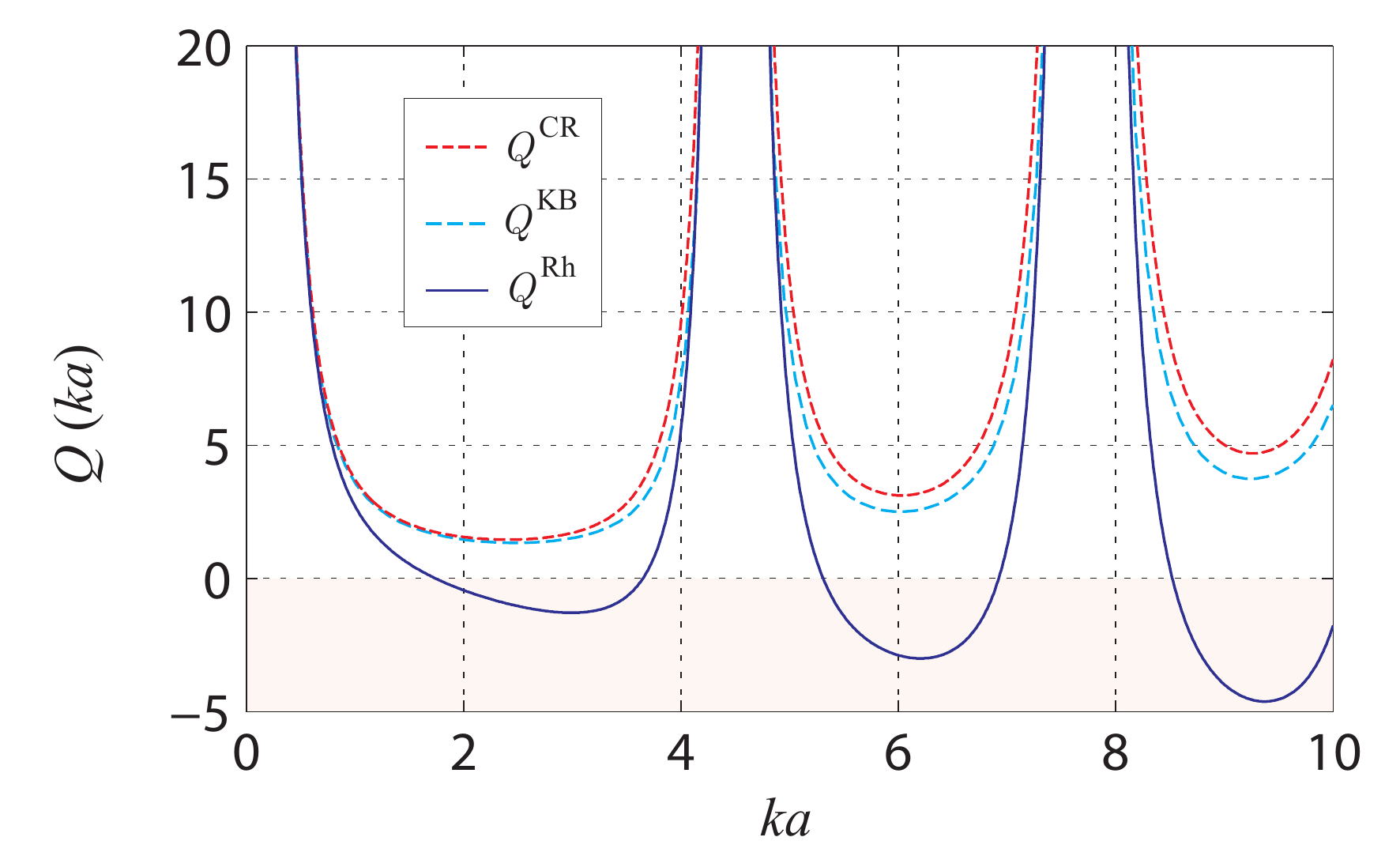}
\caption{The normalized total energy and the normalized stored energy as a function of $ka$. Several approaches to obtain the stored energy are depicted.}
\label{fig_figure3}
\end{figure}

Similarly, we can evaluate the stored energy and its radial density within the scheme of Collin-Rothschild, Kaiser-Bateman and Rhodes as
\BE
\label{Eq17}
Q^\mathrm{CR / KB / Rh} = Q_\mathrm{int}^\mathrm{CR / KB / Rh} + Q_\mathrm{ext}^\mathrm{CR / KB / Rh}
\EE
with
\begin{subequations}
\begin{align}
\label{Eq18A}
Q_\mathrm{int}^\mathrm{CR / KB / Rh} = \int\limits_0^{ka} \mathcal{Q}_\mathrm{int}^\mathrm{CR / KB / Rh} \left( kr \right) \D{kr}, \\
\label{Eq18B}
Q_\mathrm{ext}^\mathrm{CR / KB / Rh} = \int\limits_{ka}^\infty \mathcal{Q}_\mathrm{ext}^\mathrm{CR / KB / Rh} \left( kr \right) \D{kr}.
\end{align}
\end{subequations}
The densities $\mathcal{Q}_\mathrm{int}^\mathrm{tot} \left(kr\right)$, $\mathcal{Q}_\mathrm{ext}^\mathrm{tot} \left(kr\right)$, $\mathcal{Q}_\mathrm{int}^\mathrm{CR / KB / Rh} \left(kr\right)$, and $\mathcal{Q}_\mathrm{ext}^\mathrm{CR / KB / Rh} \left(kr\right)$ are depicted for $ka = 6$ in Fig.~\ref{fig_figure2}, while the $Q^\mathrm{tot}$ and $Q^\mathrm{CR / KB / Rh}$ are depicted in Fig.~\ref{fig_figure3} as functions of $ka$.


\section{Discussion}
\label{Sec_Disc}

There are several important observations in Table~\ref{ref_Table1}, Fig.~\ref{fig_figure2} and Fig.~\ref{fig_figure3} that will be discussed separately in the following subsections.

\subsection{Over-subtraction}
\label{Sec_Inside}

The curves in Fig.~\ref{fig_figure2} reveal that radiation energy subtraction inside the circumscribing sphere brings serious issues, the significance of which will grow with the electrical size of the radiator. 

Particularly, there is an observable difference between the stored energy density of Collin-Rothschild and that of Kaiser-Bateman. While the scheme of Collin-Rothschild does not subtract any radiation energy in the internal region (the power flow in the radial direction is strictly zero), the scheme of Kaiser-Bateman subtracts the entire Poynting's vector there, which is indeed non-zero also inside the sphere. A comparison of the stored energy density in Kaiser-Bateman's scheme with the total energy density reveals that there is an energy loss in the internal region provided by angular components of the Poynting's vector. This energy is however not radiated out of the sphere. This subtraction is clearly incorrect, as it will also be performed in the case of a spherical cavity, which does not contain radiation, at least not in the classical sense of an energy reaching infinity.

A different form of over-subtraction also burdens the scheme of Rhodes, but some care should be taken with the different interpretation of this scheme by various authors, namely, Rhodes \cite{Rhodes_ObservableStoredEnergiesOfElectromagneticSystems}, Yaghjian and Best \cite{YaghjianBest_ImpedanceBandwidthAndQOfAntennas}, Vandenbosch \cite{Vandenbosch_ReactiveEnergiesImpedanceAndQFactorOfRadiatingStructures}, and Gustafsson \cite{Gustaffson_StoredElectromagneticEnergy_PIER}. In particular, Rhodes \cite{Rhodes_ObservableStoredEnergiesOfElectromagneticSystems} divides the entire space similarly as in this paper, see (\ref{Eq17}), thus obtaining the internal stored energy and the external stored energy. Radiation subtraction according to the third row of Table~\ref{ref_Table1} is used only in the external region, making this scheme identical to the scheme of Collin-Rothschild \cite{CollinRotchild_EvaluationOfAntennaQ}. By contrast, Gustafsson's approach \cite{Gustaffson_StoredElectromagneticEnergy_PIER} uses subtraction everywhere, but adds $ka$ to the resulting stored energy, which for spherical radiators gives the same result as the method of Collin-Rothschild, thought it generally differs. Finally, the approach of Yaghjian and Best \cite{YaghjianBest_ImpedanceBandwidthAndQOfAntennas} and the approach of Vandebosch \cite{Vandenbosch_ReactiveEnergiesImpedanceAndQFactorOfRadiatingStructures} use subtraction everywhere, with no further compensation. This leads to over-subtraction of the order $ka$ in normalized scale that is used.

\subsection{Positive semi-definiteness}
\label{Sec_negative}

Figure~\ref{fig_figure2} and Fig.~\ref{fig_figure3} show that the stored energy density and the total stored energy are both positively semi-definite within the scheme of Collin-Rothschild and Kaiser-Bateman. In fact, the scheme of Kaiser-Bateman is manifestly positively semi-definite \cite{Kaiser_ElectromagneticInertiaReactiveEnergyAndEnergy}, and general positive semi-definiteness can also be expected from the scheme of Collin-Rothschild \cite{Collin_MinimumQofSmallAntennas}. 

By contrast, the scheme of Rhodes within the paradigm of Yaghjian, Best and Vandebosch can lead to negative stored energies, which is clearly unphysical, see also \cite{GustafssonCismasuJonsson_PhysicalBoundsAndOptimalCurrentsOnAntennas_TAP}. This issue however comes as no surprise when we realize that for a current distribution independent of frequency (the case presented in this paper), the aforementioned stored energy can be expressed as \cite{CapekJelinekHazdraEichler_MeasurableQ}
\BE
\label{Eq19}
W_\mathrm{sto}^{\mathrm{Rh}} = -\frac{1}{4} \frac{\partial}{\partial\omega} \Im\left\{\int\limits_V \V{E} \cdot \V{J}^\ast \D{V} \right\} \propto \frac{\partial X \left(\omega\right)}{\partial \omega} \left| I_0 \right|^2,
\EE
where $X$ is the reactance seen by the sources of the field, where $I_0$ corresponds to the appropriate current normalization, and where $\partial X / \partial \omega$ is well known to reach negative values in radiating systems \cite{Best_TheFosterReactanceTheoremAndQualityFactorForAntennas}. The source of this negativity problem is clearly over-subtraction of the radiation energy, see Fig.~\ref{fig_figure2}.

\subsection{Coordinate dependence}
\label{Sec_coord}
One of the basic requirements on a valid physical quantity is its independence from an absolute coordinate system. Unfortunately, this is not satisfied in the case of the first and the third line of Table~\ref{ref_Table1}, where the radiation energy definition explicitly refers to absolute coordinates. A change of the coordinate origin then leads to a change in the value of the radiated and stored energy, which contradicts the idea of energy storage being a property of the radiator. This problem of the scheme of Rhodes is in fact well known, and some ways of minimizing it have already been published \cite{YaghjianBest_ImpedanceBandwidthAndQOfAntennas,Gustaffson_StoredElectromagneticEnergy_PIER}

\section{Conclusion}
\label{Sec_Concl}
This communication has reviewed and discussed three up-to-date concepts of stored electromagnetic energy density. It has been shown on a particular example that, although sound, all three concepts yield results that to a certain degree contradict physical reality. It has been shown that the problems result from an improper definition of radiated energy density. Particularly, the concepts that have been discussed failed at least in some of the following prerequisites for a physically meaningful definition, which should
\begin{itemize}
	\item be strictly local,
	\item be coordinate independent,
	\item be gauge invariant,
	\item give a positively semi-definite energy density,
	\item give a zero value everywhere in a closed cavity.
\end{itemize}
As a result, the correct definition of radiated energy density remains an open question, despite its long history and despite being a fundamental question in the classical theory of electrodynamics.


\ifCLASSOPTIONcaptionsoff
  \newpage
\fi

\bibliographystyle{IEEEtran}
\bibliography{references_LIST}
\end{document}